%% file: main.tex
\documentclass[journal]{IEEEtran}
\usepackage{cite}
\usepackage{amsmath,amssymb,amsfonts}
\usepackage{algorithm,algpseudocode}
\usepackage{graphicx,color}
\usepackage{textcomp}
\usepackage{url}
\usepackage{amsthm}
\usepackage{pdfpages}
\newtheorem{theorem}{Theorem}
\newtheorem{proposition}{Proposition}
\newtheorem{lemma}{Lemma}
\newtheorem{corollary}{Corollary}
\newtheorem{assumption}{Assumption}

\begin{document}

\title{Coverage-Aware Routing with Approximate Equilibrium Guarantees for Location-Diversity Vehicular Crowdsensing}

\author{Luis~G.~Jaimes,~\IEEEmembership{Member,~IEEE,}
        and~Ayesha~Dina,~\IEEEmembership{Member,~IEEE}%
\thanks{This work has been submitted to the IEEE for possible publication. Copyright may be transferred without notice, after which this version may no longer be accessible.}%
\thanks{The authors are with the Department of Computer Science, Florida Polytechnic University, Lakeland, FL 33805 USA. Corresponding author: Luis G. Jaimes (e-mail: ljaimes@floridapoly.edu).}}

\markboth{}{Jaimes and Dina: Coverage-Aware Routing for Location-Diversity Vehicular Crowdsensing}

\maketitle

\input{abstract}

\input{intro}
\input{sec_related}
\input{sec_model}
\input{sec_game}
\input{sec_routing}
\input{sec_incentives}
\input{sec_eval}
\input{sec_discussion}
\input{sec_conclusion}

\bibliographystyle{IEEEtran}
\bibliography{refs}

\clearpage
\includepdf[pages=-]{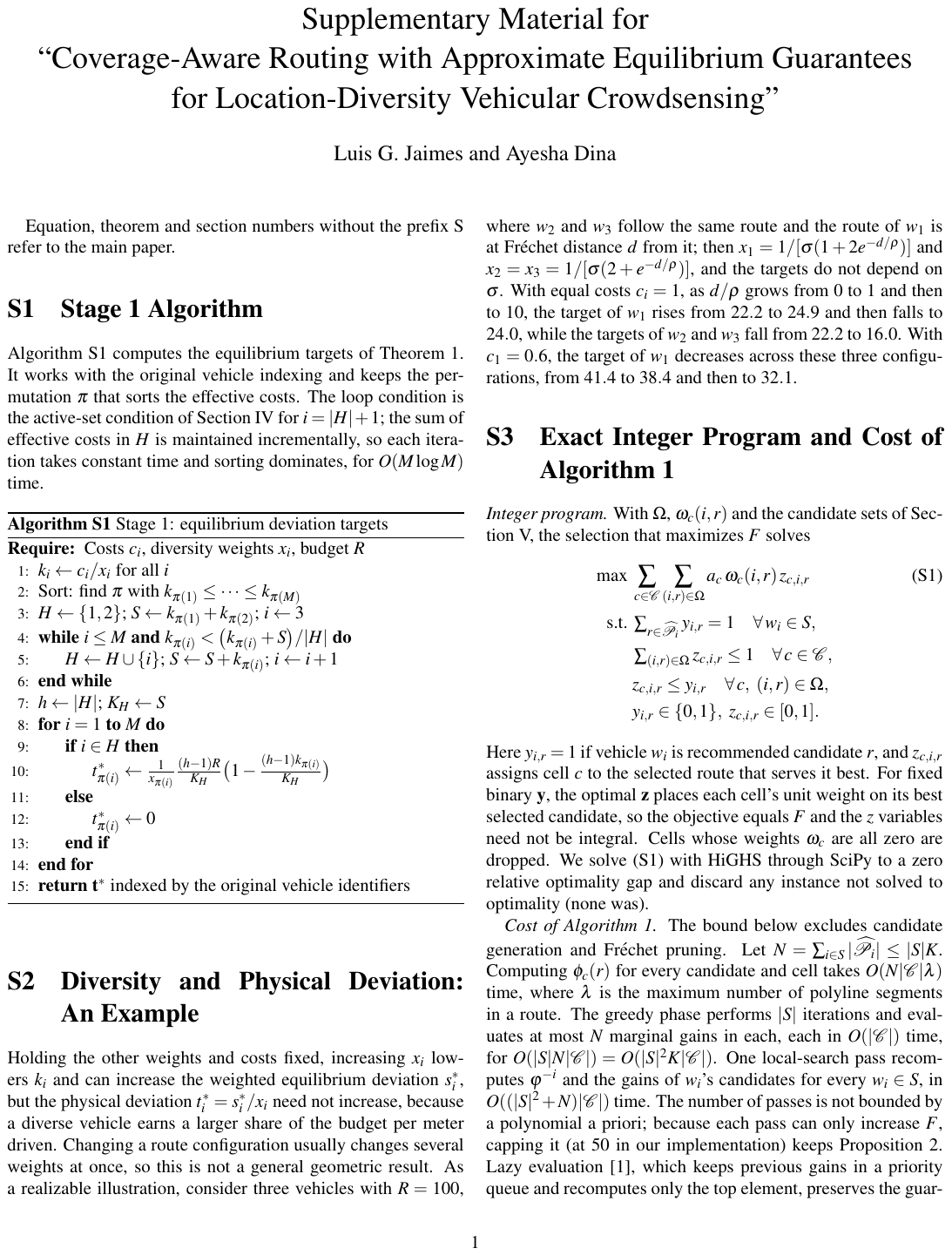}

\end{document}

%% file: abstract.tex
\begin{abstract}
Vehicular crowdsensing uses the sensors of everyday vehicles to monitor cities, but traffic concentrates on few roads, leaving much of the network unsensed. Location-diversity incentives pay vehicles more to travel where few others go, but our earlier mechanism sets how far each vehicle deviates without optimizing collective coverage in route selection. We add this step in two stages. First, a non-cooperative game with diversity-weighted rewards yields each vehicle's equilibrium deviation. Second, the platform recommends to each vehicle a feasible route whose deviation is within a tolerance of this value, chosen to improve coverage. The coverage gain is monotone submodular, so a greedy router with local search attains at least half of the optimal gain over generated candidates. If every active vehicle has a candidate within a tolerance that is small relative to the aggregate deviation, the recommended routes form an $\varepsilon$-Nash equilibrium, with $\varepsilon$ first order in the tolerance, or second order under a stronger condition. On London and Monaco road networks, the router achieves 99.4\% and 98.9\% of the optimal reduction in the coverage penalty on small instances and exceeds closest-length selection by up to 27.9 points. Under heterogeneous costs, uniform deviation needs compensation averaging 42\% of the budget to give every vehicle nonnegative utility; the equilibrium targets need almost none. They also yield more coverage per unit of added travel in most settings and up to 22.6 more points at small budgets, an advantage that depends on placement and reverses at large budgets, where detour limits bind.
\end{abstract}


%% file: intro.tex
\section{Introduction}
\label{sec:intro}

Cameras, GPS receivers and other sensors are standard on modern cars, and vehicular crowdsensing (VCS) recruits ordinary vehicles to collect data during trips they make anyway, for applications such as road surface monitoring \cite{elwakeel2018}, travel speed estimation \cite{wang2016speed}, map updating \cite{sheikh2014map} and environmental monitoring. Relying on existing trips also limits what VCS can observe: navigation systems concentrate traffic on highways and a few arterial roads, while residential streets and peripheral areas see few vehicles \cite{he2015}. Recruitment schemes that select the most useful trajectories \cite{he2015,wang2017,kong2017,chen2017} make better use of a fixed set of trips but cannot reach places no trip passes through. Reaching them requires vehicles to change routes, and the additional travel must be compensated.

Route-recommendation schemes steer vehicles toward poorly covered roads \cite{masutani2015,asprone2021,guo2023rerouting,xu2020ilocus,dimartino2022routr}, and incentive mechanisms pay vehicles to deviate \cite{chakeri2021,chakeri2022market,chintakunta2021,jaimes2024sensenow}. In our earlier work \cite{iccve2022}, we rewarded vehicles according to the location diversity of their routes, measured with the Fr\'echet distance to the other vehicles' routes, and modeled the interaction as a non-cooperative game whose unique Nash equilibrium specifies how far each vehicle should deviate. Simulations on a London road network showed higher coverage than uniform deviation.

Our earlier mechanism accounts for the spatial distribution of vehicles through these Fr\'echet-based diversity weights and uses them to compute equilibrium deviation amounts. However, it does not jointly optimize the specific routes that realize those amounts: the platform computes only the deviations, and a route of approximately the prescribed length is then chosen without considering coverage. 
This leaves two problems. First, routes satisfying the deviation targets may still provide redundant coverage, even when alternative feasible routes would improve the fleet's collective coverage. Second, the equilibrium is computed for continuous deviations, but a road network offers finitely many routes, none of which may have exactly the prescribed length, so the assigned routes need not form an equilibrium.

We address both problems with a two-stage mechanism that builds on the original incentive allocation. Stage~1 keeps the location-diversity game and computes each vehicle's equilibrium deviation. Stage~2 builds, for each vehicle, feasible routes whose deviation is within a tolerance $\delta$ of the target and within its detour limit, and selects one route per vehicle to improve the fleet's coverage. The key observation is that, with diversity weights fixed and costs proportional to added distance, routes of equal deviation give a vehicle the same utility, so the platform can choose among them freely; the incentive cost of routing comes only from the gap between realized and equilibrium deviations. The contributions are as follows.
\begin{itemize}
    \item \textbf{Two-stage formulation.} We separate how far each vehicle deviates, decided by an incentive game, from which route realizes that deviation, decided by a coverage-aware router under per-vehicle detour limits.
    \item \textbf{Coverage-aware routing.} We show that the coverage gain over candidate routes is a monotone submodular facility-location function, so a greedy router attains half of the optimal gain under the one-route-per-vehicle constraint \cite{fisher1978}; a local-search phase keeps this guarantee and closes most of the remaining gap in practice. An integer program gives exact solutions on small instances.
    \item \textbf{Incentive analysis of the routed assignment.} We prove that the recommended routes form an $\varepsilon$-Nash equilibrium of the route game whenever the aggregate weighted realization gap is at most half of the aggregate weighted equilibrium deviation, with $\varepsilon$ of first order for deviating vehicles and of second order for the others, and of second order for all vehicles under a slightly stronger condition. The bound can be computed for every instance and becomes an a priori bound in $\delta$ when every active vehicle has a candidate within the tolerance. We also give sufficient participation conditions.
    \item \textbf{Evaluation.} Using SUMO on the London network of our earlier work and the Monaco network of the MoST scenario, a factorial ablation separates the contributions of the equilibrium targets and of the router, and we test robustness to vehicle placement, grid resolution, the diversity measure and the route-generation parameters.
\end{itemize}

This article extends our conference paper \cite{iccve2022}, from which the diversity measure and the deviation game (Sections~\ref{sec:model} and~\ref{sec:game}) are summarized. The routing formulation, algorithms, incentive analysis (Sections~\ref{sec:routing} and~\ref{sec:incentives}) and the two-network evaluation (Section~\ref{sec:eval}) are new. Algorithm pseudocode, the integer program and full proofs are in the supplementary material.

%% file: sec_related.tex
\section{Related Work}
\label{sec:related}

Ji et al.~\cite{ji2023survey} survey optimization models for urban drive-by sensing. We focus on four lines of work: selecting existing trajectories, rerouting vehicles for coverage, incentives that induce deviation, and incentive design with discrete choices.

\textit{Trajectory selection with fixed trips.} Many recruitment schemes take trips as given and choose which vehicles or trajectory segments to use, maximizing coverage under a budget \cite{he2015}, minimizing cost under coverage requirements \cite{wang2017}, or acquiring segments instead of whole trips \cite{kong2017,chen2017}. Zhang et al.~\cite{zhang2015quality} show that a quality-aware coverage function is monotone submodular and obtain greedy guarantees for participant selection. None of these schemes can reach areas that no recruited trip passes through.

\textit{Rerouting vehicles for coverage.} Masutani~\cite{masutani2015} lowers link costs where sensing demand is high so that navigation routes are attracted toward it, and Asprone et al.~\cite{asprone2018,asprone2021} spread probe vehicles by selecting random routes under a constraint on path cost. Guo and Qian~\cite{guo2023rerouting} reroute unoccupied ridesourcing vehicles, selecting trip by trip, among $k$-shortest candidate routes within a travel-time window, the route that most increases the entropy of road-segment visits. Xu et al.~\cite{xu2020ilocus} plan incentives and taxi trajectories as a multiple-choice knapsack problem, and Di Martino and Starace~\cite{dimartino2022routr} propose a decentralized, budget-aware incentivized routing scheme for uniform map coverage. At the fleet level, Han et al.~\cite{han2024mixed} jointly design fleet composition, sensor allocation and routing for mixed fleets, and Dai and Han~\cite{dai2023bus} schedule buses to increase their sensing power. Our routing stage likewise selects among candidate routes between fixed origins and destinations under detour limits, but in our mechanism vehicles share a common reward in proportion to their diversity-weighted deviations, so one vehicle's route changes every other vehicle's payment. We analyze the incentive loss of mapping the continuous equilibrium onto feasible routes and give an approximation guarantee for the coverage gain.

\textit{Incentive mechanisms that induce deviation.} Many game-theoretic crowdsensing mechanisms build on the platform-centric model of Yang et al.~\cite{yang2012}, in which participants share a fixed reward in proportion to their contribution. Our group has proposed mechanisms in which autonomous vehicles choose trajectories as strategies of a non-cooperative game \cite{chakeri2021}, asynchronous vehicles account for traffic when searching for profitable routes \cite{chakeri2022market}, and vehicles are rewarded for sampling places of interest at well-spread times \cite{chintakunta2021,jaimes2024sensenow}. Our conference paper~\cite{iccve2022} rewards vehicles in proportion to their diversity-weighted deviation but does not optimize the routes that realize these deviations for collective coverage, the limitation addressed here. Wang et al.~\cite{wang2021navigation} formulate route navigation for crowdsensing as a potential game in which users share task rewards and pay detour and congestion costs, and a distributed algorithm reaches an equilibrium over routes; their model captures congestion, which our distance-based cost does not, while we obtain targets from a continuous game with a closed-form equilibrium, select routes centrally for spatial coverage, and bound profitable deviations from the routed assignment. Feng et al.~\cite{feng2025nonproprietary} combine deep reinforcement learning for route and price choice with budget-constrained task allocation weighted by willingness to traverse, and Zhang et al.~\cite{zhang2026bepim} use reference-dependent utility, loss aversion and operant conditioning to attract and retain participants in low-participation areas. Both target task completion at discrete locations rather than the effect of realizing continuous equilibrium targets as road routes.

\textit{Discrete strategies and approximation.} Ji and Chen~\cite{ji2016discretized} show that discretizing strategies, such as rounding sensing times, can invalidate contribution-allocation mechanisms, and design mechanisms for discrete strategies. In our setting the feasible deviations are the irregular, instance-dependent lengths of the available routes; rather than redesigning the mechanism for each network, we keep the continuous game and bound the incentive loss of the routed assignment in terms of the realized gaps. On the optimization side, we derive a submodular coverage gain from the capped squared-distance penalty and optimize it under a one-route-per-vehicle partition matroid \cite{fisher1978,nemhauser1978}.

%% file: sec_model.tex
\section{System Model}
\label{sec:model}

\subsection{Road Network, Vehicles, and Routes}
\label{sec:model:network}

The road network is a directed graph $G=(V,E)$ in planar coordinates measured in meters, where each edge $e$ has length $\ell_e>0$ and a polyline geometry. A \emph{route} is a simple directed path, with length $\ell(r)=\sum_{e\in r}\ell_e$; excluding routes that revisit a junction avoids paying for cyclic travel. A fleet $\mathcal{W}=\{w_1,\dots,w_M\}$ of $M\ge 2$ vehicles is available, and vehicle $w_i$ travels from $s_i$ to $d_i$ over the nonempty route set $\mathcal{R}_i$. Its default route $r_i^0$ is a shortest route, of length $L_i$, with ties broken by a fixed rule. Following route $r$, its \emph{deviation} is
\begin{equation}
t_i(r)=\ell(r)-L_i\ \ge 0 .
\label{eq:deviation}
\end{equation}
Each vehicle has a detour limit $D_i>0$ and a cost $c_i>0$ per meter of deviation, and its feasible routes are
\begin{equation}
\mathcal{R}_i^{\mathrm{feas}}=\{\, r\in\mathcal{R}_i : t_i(r)\le D_i \,\},
\label{eq:feasible}
\end{equation}
which include the default route.

\subsection{Location Diversity}
\label{sec:model:diversity}

Repeated measurements on heavily traveled roads add less spatial information than measurements where few vehicles travel, so the platform rewards vehicles whose routes are far from others'. As in \cite{iccve2022}, distances between routes are Fr\'echet distances: treating routes as curves $r_i,r_j:[0,1]\to\mathbb{R}^2$,
\begin{equation}
d_F(r_i,r_j)=\inf_{\alpha,\beta}\ \max_{\tau\in[0,1]}\ \big\| r_i(\alpha(\tau))-r_j(\beta(\tau)) \big\|,
\label{eq:frechet}
\end{equation}
over continuous nondecreasing reparameterizations $\alpha,\beta$ of $[0,1]$. The location diversity of $w_i$ is
\begin{equation}
x_i=\frac{1}{\sigma\Big[\,1+\sum_{j\neq i}\exp\!\big(-d_F(r_i^0,r_j^0)/\rho\big)\Big]},
\label{eq:diversity}
\end{equation}
with a dimensionless scale $\sigma>0$ and a length $\rho>0$ in meters. The term $1$, absent in \cite{iccve2022}, includes the vehicle's own route in the sum and bounds the weights by $1/\sigma$; because it also changes the relative weights, Section~\ref{sec:eval} evaluates its effect. The weights are computed on the default routes and announced before any route changes, so they stay fixed when vehicles deviate, which preserves the structure of the game in Section~\ref{sec:game}.

\subsection{Spatial Coverage}
\label{sec:model:coverage}

Let $\mathcal{C}$ be a partition of the area into square cells with areas $a_c$ and centers $\mathbf{p}_c$. Following \cite{iccve2022}, we use the squared distance from each cell to the nearest route as a proxy for coverage, which emphasizes areas far from every route. The \emph{coverage penalty} of a route set $\mathbf{r}$ is
\begin{equation}
J(\mathbf{r})=\sum_{c\in\mathcal{C}} a_c\, \Big(\min_{i}\ \min_{\mathbf{q}\in r_i}\ \|\mathbf{p}_c-\mathbf{q}\|\Big)^2 ,
\label{eq:penalty}
\end{equation}
with distances measured to the routes' polylines. Lower $J$ means better coverage; the coverage of \cite{iccve2022} is $U=1/J$, and since a percentage reduction in $J$ differs from a percentage increase in $U$, we report reductions in $J$.

\subsection{Participation, Information, and Payments}
\label{sec:model:platform}

All vehicles in the fleet provide sensing data along their trips, and the mechanism compensates additional travel; a vehicle may decline a detour and keep its default route. The platform knows each vehicle's origin, destination, cost and detour limit, that is, we assume they are reported truthfully, and it can verify the route each vehicle drives, for example from GPS traces. With a budget $R>0$, a vehicle whose driven route has deviation $t_i$ receives a share of $R$ proportional to its diversity-weighted deviation,
\begin{equation}
P_i(\mathbf{t})=
\begin{cases}
\dfrac{x_i t_i}{\sum_{j} x_j t_j}\, R, & \text{if } \sum_j x_j t_j>0,\\[2mm]
0, & \text{otherwise,}
\end{cases}
\label{eq:payment}
\end{equation}
where $\mathbf{t}$ are the \emph{realized} deviations, and its utility is
\begin{equation}
u_i(\mathbf{t})=P_i(\mathbf{t})-c_i t_i .
\label{eq:utility}
\end{equation}
Paying on realized deviation keeps the utility of \cite{iccve2022} and makes payments depend only on what each vehicle drives. Because the weights are fixed and costs are proportional to added distance, $u_i$ depends on $w_i$'s route only through its deviation, for any fixed choices of the others; Section~\ref{sec:incentives} builds on this property.

The mechanism proceeds as follows: (i)~the platform announces $R$ and vehicles report their trips, costs and limits; (ii)~the platform computes default routes and diversity weights; (iii)~Stage~1 computes equilibrium deviation targets $\mathbf{t}^*$ of the continuous game defined by \eqref{eq:utility} (Section~\ref{sec:game}), which need not respect the detour limits or be realizable; (iv)~Stage~2 recommends a feasible route to each vehicle (Section~\ref{sec:routing}), each vehicle drives the recommended route or another feasible route, and payments follow \eqref{eq:payment}. We assume $M\ge2$, $R>0$, and positive finite costs and weights; with a single vehicle, any positive deviation earns the whole budget and the game has no pure equilibrium.

%% file: sec_game.tex
\section{Stage 1: Deviation Targets from the Location-Diversity Game}
\label{sec:game}

Stage~1 solves the continuous relaxation of the game defined by \eqref{eq:utility}, in which each vehicle chooses a deviation $t_i\ge 0$ without regard to the road network or its detour limit. The solution follows from our earlier work \cite{iccve2022} and the sensing game of Yang et al.~\cite{yang2012}; we restate it and derive the properties used later. The resulting deviations are \emph{targets}, which Stage~2 realizes with routes.

With the \emph{weighted deviation} $s_i=x_i t_i$ and the \emph{effective cost} $k_i=c_i/x_i$, the utility becomes
\begin{equation}
u_i=\frac{s_i}{\sum_{j} s_j}\,R-k_i s_i
\label{eq:yanggame}
\end{equation}
whenever $\sum_j s_j>0$, and $u_i=0$ when all $s_j=0$. This is the platform-centric sensing game of \cite{yang2012}, with sensing times replaced by weighted deviations and unit costs by effective costs. Because $t_i\mapsto x_it_i$ is a bijection of $[0,\infty)$ that preserves utilities and unilateral deviations, the equilibria of the two games correspond one to one.

Relabel the vehicles in order of increasing effective cost, $k_1\le\dots\le k_M$, with ties broken by a fixed rule. The \emph{active set} is $H=\{1,\dots,h\}$, where $h\ge 2$ is the largest index such that
\begin{equation}
k_i<\frac{1}{i-1}\sum_{j=1}^{i}k_j \qquad \text{for all } i=2,\dots,h ,
\label{eq:activeset}
\end{equation}
which always holds for $i=2$. Let $K_H=\sum_{j\in H}k_j$ and $a_i=(h-1)k_i/K_H$.

\begin{theorem}[\cite{yang2012,iccve2022}]
\label{thm:stage1}
Under the assumptions of Section~\ref{sec:model}, the continuous deviation game has a unique Nash equilibrium,
\begin{equation}
t_i^*=
\begin{cases}
\dfrac{(h-1)R}{x_iK_H}\,(1-a_i), & i\in H,\\[2mm]
0, & i\notin H.
\end{cases}
\label{eq:tstar}
\end{equation}
\end{theorem}

\begin{proof}
Yang et al.\ show that the transformed game \eqref{eq:yanggame} has the unique equilibrium $s_i^*=\frac{(h-1)R}{K_H}(1-a_i)$ for $i\in H$ and $s_i^*=0$ otherwise \cite[Ths.~1 and~2]{yang2012}; dividing by $x_i$ gives \eqref{eq:tstar}.
\end{proof}

The targets are computed by sorting the effective costs and adding vehicles to $H$ until \eqref{eq:activeset} first fails, in $O(M\log M)$ time; the pseudocode is in the supplementary material.

\begin{proposition}
\label{prop:stage1}
Fix the costs and the diversity weights. (a)~The active set $H$ does not depend on $R$, and $\mathbf{t}^*(R)=R\,\mathbf{t}^*(1)$. (b)~Every $i\in H$ has equilibrium utility
\begin{equation}
u_i(\mathbf{t}^*)=R(1-a_i)^{2}>0 ,
\label{eq:ustar}
\end{equation}
and $u_i(\mathbf{t}^*)=0$ for $i\notin H$. (c)~The aggregate weighted deviation is $S^*=\sum_j x_jt_j^*=(h-1)R/K_H$, and the full budget is paid.
\end{proposition}

\begin{proof}
(a) $H$ depends only on the effective costs, and \eqref{eq:tstar} is linear in $R$. (c) Summing $s_i^*$ over $H$ gives $\frac{(h-1)R}{K_H}(h-(h-1))$. (b) The payment share of $i\in H$ is $s_i^*/S^*=1-a_i$ and its cost is $k_is_i^*=Ra_i(1-a_i)$, so $u_i=R(1-a_i)^2$; condition \eqref{eq:activeset} with $i=h$ gives $a_h<1$, and $a_i\le a_h$.
\end{proof}

Property~(a) means that the budget only scales the targets; it does not change which vehicles deviate. Property~(b) gives every active vehicle a margin $R(1-a_i)^2$ over staying on its default route, which the participation conditions of Section~\ref{sec:incentives} use. Diversity enters only through the effective costs: a larger $x_i$ lowers $k_i$ and can raise the weighted deviation $s_i^*$, but not necessarily the physical deviation $t_i^*=s_i^*/x_i$, so the mechanism differentiates incentives by diversity without guaranteeing that the most isolated vehicles deviate most (an example is in the supplementary material). Inactive vehicles have target zero and are recommended their default routes, but they remain strategic players, and Section~\ref{sec:incentives} covers them as well.

%% file: sec_routing.tex
\section{Stage 2: Coverage-Aware Route Selection}
\label{sec:routing}

Stage~2 turns the targets $\mathbf{t}^*$ into route recommendations: it builds a finite set of feasible candidate routes near each active vehicle's target and selects one candidate per vehicle so that the selected routes jointly reduce the coverage penalty. It does not depend on how the targets were computed, so the same router also serves the uniform targets of Section~\ref{sec:eval}.

\subsection{Candidate Routes}
\label{sec:routing:candidates}

Let $A$ be the active vehicles of Stage~1. For a tolerance $\delta\ge0$ in meters, the routes that realize the target of $w_i\in A$ are
\begin{equation}
\mathcal{P}_i(\delta)=\{\, r\in\mathcal{R}_i^{\mathrm{feas}} : |t_i(r)-t_i^*|\le\delta \,\}.
\label{eq:candidates}
\end{equation}
Since enumerating $\mathcal{P}_i(\delta)$ is impractical, Stage~2 generates a subset $\widehat{\mathcal{P}}_i\subseteq\mathcal{P}_i(\delta)$ of at most $K$ routes. It first builds a pool $\mathcal{Q}_i$ of simple, detour-feasible \emph{via-routes}, each a shortest path from $s_i$ to a junction $v$ followed by a shortest path from $v$ to $d_i$; one forward and one backward shortest-path computation give the lengths of all via-routes, and the pool always contains the default route. It then keeps the routes whose deviation is within $\delta$ of the target and, in order of increasing realization gap $|t_i(r)-t_i^*|$, retains a route unless its Fr\'echet distance to an already retained route is below a threshold $\eta$, stopping after $K$ routes. The guarantees below hold with respect to the generated sets.

\textit{Vehicles without candidates.} If $\widehat{\mathcal{P}}_i$ is empty, because $t_i^*>D_i+\delta$, because no simple route has a deviation within the tolerance, or because the generator missed one, Stage~2 recommends the fallback route
\begin{equation}
r_i^{\mathrm{fb}}\in\arg\min_{r\in\mathcal{Q}_i}|t_i(r)-t_i^*| ,
\label{eq:fallback}
\end{equation}
and records the vehicle as \emph{unrealized}. The a priori incentive bounds of Section~\ref{sec:incentives} require the following condition; the actual-gap certificate of that section also applies to fallback cases whenever its aggregate-gap condition holds.

\begin{assumption}[Realizability within tolerance]
\label{ass:realizable}
Every active vehicle $w_i\in A$ has a generated candidate within tolerance $\delta$ of its target, that is, $\widehat{\mathcal{P}}_i\neq\emptyset$.
\end{assumption}

Inactive vehicles (recommended their default routes) and unrealized vehicles (recommended $r_i^{\mathrm{fb}}$) have recommendations that are fixed during coverage optimization, although they remain strategic players. Let $\mathbf{b}$ be the set of these fixed routes, possibly empty, and $S\subseteq A$ the active vehicles with candidates; the default routes of vehicles in $S$ are not in $\mathbf{b}$, because they are replaced.

\subsection{Coverage Objective and Submodularity}
\label{sec:routing:objective}

For a cell $c$ and a route $r$, let $\phi_c(r)=\min\{\bar d^{\,2},\min_{\mathbf{q}\in r}\|\mathbf{p}_c-\mathbf{q}\|^2\}$ be the squared distance capped at $\bar d^{\,2}$, and $\phi_c(\mathbf{b})=\min_{r\in\mathbf{b}}\phi_c(r)$ (or $\bar d^{\,2}$ if $\mathbf{b}=\emptyset$). Selecting $r_i\in\widehat{\mathcal{P}}_i$ for each $w_i\in S$ yields the \emph{coverage gain}
\begin{equation}
F(\mathbf{r}_S)=\sum_{c\in\mathcal{C}} a_c\Big[\phi_c(\mathbf{b})-\min\big\{\phi_c(\mathbf{b}),\ \min_{i\in S}\phi_c(r_i)\big\}\Big] ,
\label{eq:gain}
\end{equation}
the reduction in the capped penalty over the fixed routes. With the reference penalty $B=\sum_c a_c\phi_c(\mathbf{b})$, if the cap never truncates a distance, $F(\mathbf{r}_S)=B-J(\mathbf{b}\cup\mathbf{r}_S)$, so maximizing $F$ minimizes the penalty \eqref{eq:penalty} of the completed fleet. A ratio on $F$ is therefore a ratio on the reduction from $B$: $F_{\mathrm{greedy}}\ge\frac12F_{\mathrm{opt}}$ implies $J_{\mathrm{greedy}}\le\frac12(B+J_{\mathrm{opt}})$, which is weak when $B$ is large, for example when $\mathbf{b}=\emptyset$ and $B=\bar d^{\,2}\sum_ca_c$. We therefore emphasize reductions in $J$ relative to shortest routes and report $F/F_{\mathrm{opt}}$ separately to assess the approximation guarantee.

Let $\Omega=\{(i,r): w_i\in S,\ r\in\widehat{\mathcal{P}}_i\}$ and $\omega_c(i,r)=\max\{0,\phi_c(\mathbf{b})-\phi_c(r)\}$. Extending \eqref{eq:gain} to $X\subseteq\Omega$ by $F(X)=\sum_c a_c\max_{(i,r)\in X}\omega_c(i,r)$, with $F(\emptyset)=0$, the selections of one candidate per vehicle are the bases of the partition matroid $\mathcal{M}=\{X\subseteq\Omega:|X\cap\Omega_i|\le1\ \forall i\}$, where $\Omega_i$ holds the pairs of $w_i$.

\begin{lemma}
\label{lem:submodular}
$F$ is nonnegative, monotone and submodular on $\Omega$.
\end{lemma}

\begin{proof}
$F$ is a facility-location function with nonnegative weights, which is monotone and submodular \cite{nemhauser1978}: adding an element can only raise each per-cell maximum, by no more than it would for a smaller set, and nonnegative sums keep both properties.
\end{proof}

\subsection{Greedy Routing with Local Search}
\label{sec:routing:greedy}

For per-cell penalties $\varphi=(\varphi_c)$, let $g_{\varphi}(r)=\sum_{c}a_c\max\{0,\varphi_c-\phi_c(r)\}$ be the gain of route $r$ over $\varphi$. Algorithm~\ref{alg:greedy} first assigns routes greedily by largest marginal gain, breaking ties, including zero gains, by the smallest realization gap. Because the greedy phase never revisits a choice, an early route can become redundant, so a local-search phase then gives each vehicle in turn its best candidate given the others' routes, replacing a route only if the gain strictly increases, until no route changes.

\begin{algorithm}[t]
\caption{Stage 2: greedy coverage-aware routing with local search}
\label{alg:greedy}
\begin{algorithmic}[1]
\Require Fixed routes $\mathbf{b}$; candidate sets $\widehat{\mathcal{P}}_i$ for $w_i\in S$; targets $t_i^*$
\State $\varphi_c\gets\phi_c(\mathbf{b})$ for all $c\in\mathcal{C}$; $Q\gets S$ \Comment{greedy phase}
\While{$Q\neq\emptyset$}
    \State $(i,r)\gets\arg\max_{w_i\in Q,\ r\in\widehat{\mathcal{P}}_i} g_{\varphi}(r)$ \Comment{ties: smallest gap}
    \State $r_i\gets r$; $Q\gets Q\setminus\{w_i\}$; $\varphi_c\gets\min\{\varphi_c,\phi_c(r)\}\ \forall c$
\EndWhile
\Repeat \Comment{local-search phase}
    \State $\mathit{changed}\gets\mathbf{false}$
    \For{$w_i\in S$ in a fixed order}
        \State $\varphi^{-i}_c\gets\min\{\phi_c(\mathbf{b}),\ \min_{j\in S\setminus\{i\}}\phi_c(r_j)\}$ for all $c$
        \State $r'\gets\arg\max_{r\in\widehat{\mathcal{P}}_i} g_{\varphi^{-i}}(r)$ \Comment{ties: smallest gap}
        \If{$g_{\varphi^{-i}}(r')>g_{\varphi^{-i}}(r_i)$}
            \State $r_i\gets r'$; $\mathit{changed}\gets\mathbf{true}$
        \EndIf
    \EndFor
\Until{\textbf{not} $\mathit{changed}$}
\State \Return routes $r_i$ and realization gaps $|t_i(r_i)-t_i^*|$
\end{algorithmic}
\end{algorithm}

\begin{proposition}
\label{prop:greedy}
Algorithm~\ref{alg:greedy} terminates and returns a selection whose value of $F$ is at least half the maximum over all selections of one generated candidate per vehicle in $S$; the same holds if the local-search phase is stopped after any number of passes.
\end{proposition}

\begin{proof}
By Lemma~\ref{lem:submodular}, the greedy phase achieves at least half the optimum under the partition matroid \cite{fisher1978}; tie-breaking does not affect the argument. In the local-search phase, $F$ equals $g_{\varphi^{-i}}(r_i)$ plus a term independent of $w_i$'s route, so each replacement strictly increases $F$; hence $F$ never decreases after the greedy phase, no selection repeats, and the phase terminates.
\end{proof}

The continuous greedy algorithm improves the ratio to $1-1/e$ \cite{calinescu2011} at a higher cost. The guarantee concerns $F$, which credits all the coverage of the selected routes beyond $\mathbf{b}$, and says little about the reduction in $J$ relative to shortest routes, which can be a small part of $F$; Section~\ref{sec:eval} reports both, together with the \emph{greedy-only} rule that stops after the greedy phase. With $N=\sum_{i\in S}|\widehat{\mathcal{P}}_i|\le|S|K$ candidates, the greedy phase takes $O(|S|N|\mathcal{C}|)$ time and one local-search pass $O((|S|^2+N)|\mathcal{C}|)$; our implementation caps local search at 50 passes, which keeps the guarantee, and lazy evaluation \cite{minoux1978} can accelerate the greedy phase. The supplementary material gives the detailed cost analysis.

\textit{Exact solution and baselines.} To measure the gap of Algorithm~\ref{alg:greedy}, we solve the selection exactly on small instances with a facility-location integer program that chooses one candidate per vehicle and assigns each cell to the selected route that serves it best; its objective equals $F$, it has $N$ binary and $N|\mathcal{C}|$ continuous variables, and its formulation is in the supplementary material. Two controlled baselines use the same candidate sets: \emph{closest-length selection}, which picks the candidate with the smallest realization gap, as our earlier implementation did, and \emph{random selection}, as described in \cite{iccve2022}.

%% file: sec_incentives.tex
\section{Incentive Analysis of the Routed Assignment}
\label{sec:incentives}

Stage~1 computes an equilibrium of the continuous relaxation, but Stage~2 recommends routes whose deviations differ from the targets. This section bounds how much any vehicle, active or inactive, can gain by ignoring its recommendation, and gives conditions under which active vehicles still prefer to deviate.

\subsection{The Route Game and the Best-Response Value}
\label{sec:incentives:game}

In the route game, the strategies of vehicle $w_i$ are its feasible routes $\mathcal{R}_i^{\mathrm{feas}}$, including the default route, and its payoff is $u_i(\mathbf{t}(\mathbf{r}))$ with $u_i$ given by \eqref{eq:utility}. A route profile $\hat{\mathbf{r}}$ is an \emph{$\boldsymbol{\varepsilon}$-Nash equilibrium}, $\boldsymbol{\varepsilon}\ge0$, if
\begin{equation}
u_i\big(\mathbf{t}(r_i,\hat{\mathbf{r}}_{-i})\big)\le u_i\big(\mathbf{t}(\hat{\mathbf{r}})\big)+\varepsilon_i
\qquad \forall\, r_i\in\mathcal{R}_i^{\mathrm{feas}},\ \forall\, i .
\label{eq:epsne}
\end{equation}
For the profile $\hat{\mathbf{r}}$ recommended by Stage~2, with deviations $\hat t_i=t_i(\hat r_i)$, the \emph{realization gap} of $w_i$ is $e_i=|\hat t_i-t_i^*|$, which is zero for inactive vehicles, and the aggregate weighted gap is
\begin{equation}
\Delta=\sum_{j}x_je_j .
\label{eq:Delta}
\end{equation}
We keep the sorted indexing and notation of Section~\ref{sec:game}, with $S^*=(h-1)R/K_H$, and write $A_i^*=S^*-s_i^*$ and $\hat A_i$ for the aggregate weighted deviation of the vehicles other than $w_i$ at the targets and at the recommendation; $A_i^*=a_iS^*$ for $i\in H$.

For a vehicle with effective cost $k$ facing an aggregate $A$ from the others, let $f(s,A)=\frac{s}{A+s}R-ks$ for $s>0$, $f(0,A)=0$, and $V(A)=\sup_{s\ge 0}f(s,A)$.

\begin{lemma}
\label{lem:brvalue}
For $A>0$, $V(A)=(\sqrt{R}-\sqrt{kA})^2$ if $kA<R$, and $V(A)=0$ otherwise; $V(0)=R$ is a supremum that no positive deviation attains.
\end{lemma}

\begin{proof}
For $A>0$, $f(\cdot,A)$ is concave with $\partial f/\partial s=RA/(A+s)^2-k$. If $kA\ge R$ the supremum is $f(0,A)=0$; otherwise the maximizer $s=\sqrt{RA/k}-A$ gives payment share $1-\sqrt{kA/R}$ and cost $\sqrt{kAR}-kA$. For $A=0$, $f(s,0)=R-ks$.
\end{proof}

Every feasible route has a deviation $t_i(r_i)\ge0$, so $V$ evaluated at $\hat A_i$, with the vehicle's own effective cost, bounds the utility of any unilateral route change, whether or not the route is a candidate and regardless of the detour limit.

\subsection{Approximate Equilibrium}
\label{sec:incentives:theorem}

\begin{theorem}
\label{thm:epsne}
If $\Delta\le S^*/2$, the recommended profile $\hat{\mathbf{r}}$ is an $\boldsymbol{\varepsilon}$-Nash equilibrium of the route game with
\begin{align}
\varepsilon_i&=R\left(\frac{\Delta}{S^*}\right)^{2}, && i\notin H, \label{eq:epsinactive}\\
\varepsilon_i&=\frac{R}{S^*}\left(\frac{\Delta}{a_i}+2\Delta+2x_ie_i\right)+c_ie_i, && i\in H. \label{eq:epsactive}
\end{align}
\end{theorem}

\begin{proof}[Proof sketch]
It suffices to bound $V(\hat A_i)-u_i(\hat{\mathbf{t}})$. The condition gives $|\hat A_i-A_i^*|\le\Delta$ and a realized aggregate of at least $S^*/2$. An inactive vehicle has effective cost at least $R/S^*$, because the active-set rule fails at position $h+1$, so Lemma~\ref{lem:brvalue} gives $V(\hat A_i)\le R(1-\sqrt{1-\Delta/S^*})^2\le R(\Delta/S^*)^2$. For an active vehicle, $u_i(\mathbf{t}^*)=V(A_i^*)$; the change of $V$ from $A_i^*$ to $\hat A_i$ is at most $R\Delta/A_i^*$, and the change of utility from the targets to the recommendation is bounded by the partial derivatives of $f$ along the segment between them, which gives \eqref{eq:epsactive}. The full proof is in the supplementary material.
\end{proof}

The bound for inactive vehicles is of second order in the gaps, and the bound for active vehicles of first order; its term $\Delta/a_i$ is largest for the vehicle with the smallest effective cost. Theorem~\ref{thm:epsne} uses only the realized gaps, so it gives a certificate for every instance, including instances with fallback routes, whenever $\Delta\le S^*/2$; fallback gaps are not controlled by $\delta$, so the condition may then fail or the certificate may be large. Under Assumption~\ref{ass:realizable} the gaps are controlled a priori.

\begin{corollary}
\label{cor:delta}
Under Assumption~\ref{ass:realizable}, $e_i\le\delta$ and $\Delta\le\delta\sum_{j\in H}x_j$. If
\begin{equation}
\delta\le\frac{S^*}{2\sum_{j\in H}x_j},
\label{eq:deltacond}
\end{equation}
the recommended profile is an $\boldsymbol{\varepsilon}$-Nash equilibrium with \eqref{eq:epsinactive} and \eqref{eq:epsactive} evaluated at $e_i=\delta$ and $\Delta=\delta\sum_{j\in H}x_j$; hence $\max_i\varepsilon_i=O(\delta)$, and $\varepsilon_i=O(\delta^2)$ for inactive vehicles.
\end{corollary}

\textit{Second-order refinement.} The active bound is of first order because it ignores that each active vehicle is at its best response at the targets, where its utility is flat. Under a slightly stronger condition the loss is of second order.

\begin{proposition}
\label{prop:second}
Suppose $\Delta\le S^*/2$, and let $i\in H$ satisfy
\begin{equation}
\Delta\le\frac12\min\Big\{1,\ \frac{1-a_i^2}{a_i^2}\Big\}A_i^* .
\label{eq:secondcond}
\end{equation}
Then no route change can increase the utility of $w_i$ by more than
\begin{equation}
\varepsilon_i^{(2)}=\frac{R}{T_i^2}\big(x_ie_i+\Lambda_i\Delta\big)^2\ \le\ \frac{4R}{(S^*)^2}\big(x_ie_i+\Lambda_i\Delta\big)^2,
\label{eq:epssecond}
\end{equation}
where $T_i=\min\{\hat S,\sqrt{R\hat A_i/k_i}\}\ge S^*/2$, $\hat S$ is the realized aggregate, and $\Lambda_i=\max\{1,1/(\sqrt2\,a_i)\}$.
\end{proposition}

\begin{proof}[Proof sketch]
Condition \eqref{eq:secondcond} keeps the best response to $\hat A_i$ positive, so the regret equals half the curvature of $f$ times the squared distance to the best response; the curvature is at most $2R/T_i^2$, and the distance is at most $x_ie_i+\Lambda_i\Delta$ because the best response is Lipschitz with constant $\Lambda_i$ near $A_i^*$. The full proof is in the supplementary material.
\end{proof}

Under Assumption~\ref{ass:realizable}, every active vehicle that satisfies \eqref{eq:secondcond} has $\varepsilon_i^{(2)}=O(\delta^2)$, so together with \eqref{eq:epsinactive} all vehicles then have second-order losses. Like Theorem~\ref{thm:epsne}, the bound uses only realized quantities.

\subsection{Participation}
\label{sec:incentives:participation}

The default route yields zero deviation, payment and cost, so an active vehicle prefers its recommendation if its utility there is positive.

\begin{corollary}
\label{cor:ir}
Suppose $\Delta\le S^*/2$. An active vehicle $w_i$ strictly prefers its recommended route to its default route if
\begin{equation}
R\,(1-a_i)^2>\frac{2R}{S^*}\big(x_ie_i+\Delta\big)+c_ie_i .
\label{eq:ircond}
\end{equation}
Under Assumption~\ref{ass:realizable} and \eqref{eq:deltacond}, substituting $e_i=\delta$ and $\Delta=\delta\sum_{j\in H}x_j$ gives a condition that holds for all sufficiently small $\delta$ at which realizability holds.
\end{corollary}

\begin{proof}
By Proposition~\ref{prop:stage1}(b), $u_i(\mathbf{t}^*)=R(1-a_i)^2$, and the derivative bound in the proof of Theorem~\ref{thm:epsne} gives $u_i(\hat{\mathbf{t}})\ge R(1-a_i)^2-\frac{2R}{S^*}(x_ie_i+\Delta)-c_ie_i$; the right side of \eqref{eq:ircond} vanishes as $\delta\to0$, while $a_i<1$.
\end{proof}

Condition \eqref{eq:ircond} is sufficient, not necessary: the margin $R(1-a_i)^2$ is the utility that rounding may consume before a vehicle would rather stay on its default route. Inactive vehicles need no such condition, because their recommendation is the default route.

%% file: sec_eval.tex
\section{Evaluation}
\label{sec:eval}

We evaluate how close Algorithm~\ref{alg:greedy} comes to the exact optimum, how much coverage-aware selection and the equilibrium deviation targets each contribute, how the routed assignments preserve incentives, and how robust these findings are to the network, the vehicle placement and the numerical parameters.

\subsection{Setup}
\label{sec:eval:setup}

\textit{Networks.} We use two road networks in SUMO \cite{krajzewicz2012sumo}. The first is the downtown London network of \cite{iccve2022}, imported from OpenStreetMap: 14,655 junctions in an area of 8.4\,km by 6.5\,km. The second is the road network of the Monaco SUMO Traffic (MoST) scenario \cite{codeca2017most}, of which we use only the road graph and not the traffic demand: 2,004 junctions in a box of 10.0\,km by 6.4\,km, with a denser and more constrained street layout than London. Both are used with SUMO 1.28.0, and the MoST network is taken from commit \texttt{b29b2f6} of its public repository. The coverage domain $\mathcal{C}$ consists of the 100\,m cells whose center lies within 500\,m of a road open to passenger cars; this rule, fixed before the comparative results were examined, keeps 5,458 of the 5,460 cells of the London box and 3,707 of the 6,400 cells of the Monaco box, which includes sea and land without roads. Distances are computed to road polylines, and the cap $\bar d$ is the diagonal of the box, so it never truncates.

\textit{Fleet and placement.} As in \cite{iccve2022}, the main fleet has two groups that cross the area, 4 vehicles in one group and 10 in the other, so the small group is spatially isolated (\emph{clustered} placement). Origins and destinations are drawn uniformly among junctions within 1\,km of each group's reference junctions, with a straight-line separation of at least 2\,km; shortest routes average 4.6\,km in London and 5.3\,km in Monaco. To test whether the results depend on this configuration, we also use a \emph{balanced} placement with groups of 7 and 7 vehicles, and a \emph{dispersed} placement in which the 14 origins and destinations are drawn uniformly from the largest strongly connected component, with straight-line separation between 2 and 6\,km. Each experiment uses 20 scenarios drawn from fixed seeds, and every scenario is reused for every budget and method, so all comparisons are paired.

\textit{Mechanism parameters.} Budgets are $R\in\{5,15,25,35\}\times10^{3}$. Costs per meter are either equal ($c_i=1$) or heterogeneous, drawn from a lognormal distribution with mean $1$ and log-standard deviation $0.5$. Diversity uses $\sigma=1$ and $\rho=1000$\,m. Detour limits are $D_i=L_i$, with $D_i\in\{0.5L_i,1.5L_i\}$ in a sensitivity study. Stage~2 uses tolerance $\delta=100$\,m, at most $K=20$ candidates per vehicle and pruning threshold $\eta=100$\,m.

\textit{Methods and measures.} We cross two deviation rules with four selection rules. The deviation rules are LDD, the equilibrium targets of Section~\ref{sec:game}, and \emph{uniform}, which gives every vehicle the mean LDD target, so that both rules have the same total target deviation. The selection rules, applied to the same candidate sets, are closest-length, random, greedy-only (the greedy phase of Algorithm~\ref{alg:greedy}) and the full Algorithm~\ref{alg:greedy}, which we call coverage-aware routing. We report the reduction in the coverage penalty $J$ relative to shortest routes, the added travel, the number of unrealized vehicles, utilities under the payment rule \eqref{eq:payment}, and, for LDD, the certificates of Section~\ref{sec:incentives} together with each vehicle's best-response gain, the largest gain from any unilateral change of deviation $t_i\ge0$ computed exactly by Lemma~\ref{lem:brvalue}. Because it ignores which routes exist and the detour limits, it is an upper bound on the gain from any feasible route change. Intervals are 95\% $t$-intervals over scenarios, and differences between methods are paired by scenario. The implementation and all configurations, seeds and raw results are available at \url{https://github.com/ljaimesco/coverage-aware-routing-vcs}.

\subsection{Coverage-Aware Routing versus the Exact Optimum}
\label{sec:eval:ilp}

To compare Algorithm~\ref{alg:greedy} with the exact integer program (Section~\ref{sec:routing:greedy} and Supplement~S3), we use 10 smaller scenarios per network with 6 vehicles (2 and 4 per group), heterogeneous costs, $R\in\{1,3\}\times10^{3}$, $K=8$ and 400\,m cells, which gives 40 instances per network with both deviation rules. One Monaco instance in which no vehicle has a candidate to select, so that every rule gives the same outcome, is excluded, and a second, in which the optimal reduction in $J$ is not above $10^{-6}$ of $J$ for shortest routes (it is negative), is excluded from the ratio of reductions in $J$ only. The integer program is solved with HiGHS to a zero relative optimality gap, and a selection counts as optimal when its value of $F$ is within a relative $10^{-9}$ of the optimum. Table~\ref{tab:ilp} reports two ratios to the optimum: of the coverage gain $F$, which Proposition~\ref{prop:greedy} bounds, and of the reduction in $J$ relative to shortest routes, which is the quantity of practical interest. They differ because $F$ is measured from the fixed routes $\mathbf{b}$, which exclude the default routes of the vehicles being routed, so $F$ credits all the coverage of the selected routes, whereas the reduction in $J$ credits only what they add beyond the shortest routes. Mean $F/F_{\mathrm{opt}}$ ratios are high for all rules, so the reduction in $J$ is the more discriminating measure, and the guarantee on $F$ implies little about it. Algorithm~\ref{alg:greedy} attains on average 99.4\% and 98.9\% of the optimal reduction in $J$. Local search matters most in the worst case: in one Monaco instance the greedy phase alone attains only 10.2\% of the optimal reduction in $J$, which is itself only 0.33\%, while attaining 99.97\% of the optimal $F$, consistent with Proposition~\ref{prop:greedy}; local search reaches the optimum there. Closest-length and random selection sometimes increase $J$.

\begin{table}[t]
\centering
\caption{Coverage-aware routing versus the exact integer program (small instances, 6 vehicles)}
\label{tab:ilp}
\small
\setlength{\tabcolsep}{4pt}
\begin{tabular}{@{}p{0.46\columnwidth}cc@{}}
\hline
 & London & Monaco \\
\hline
Instances, $F$ ratio / $J$ ratio & 40 / 40 & 39 / 38 \\
Optimal selections, Alg.~\ref{alg:greedy} / greedy-only & 32 / 20 & 32 / 21 \\
Fraction of optimal $F$, mean (worst): & & \\
\quad Alg.~\ref{alg:greedy} & 1.000 (0.997) & 1.000 (0.996) \\
\quad greedy-only & 0.998 (0.936) & 1.000 (0.993) \\
\quad closest-length & 0.963 (0.467) & 0.998 (0.975) \\
\quad random & 0.961 (0.527) & 0.984 (0.446) \\
Fraction of optimal reduction in $J$, mean (worst): & & \\
\quad Alg.~\ref{alg:greedy} & 0.994 (0.896) & 0.989 (0.833) \\
\quad greedy-only & 0.976 (0.793) & 0.943 (0.102) \\
\quad closest-length, mean & 0.43 & 0.57 \\
\quad random, mean & 0.46 & 0.27 \\
Selection time, median: & & \\
\quad Alg.~\ref{alg:greedy} & 0.6\,ms & 0.5\,ms \\
\quad ILP & 0.28\,s & 0.02\,s \\
\hline
\end{tabular}
\end{table}

\subsection{Contribution of Routing and of the Deviation Targets}
\label{sec:eval:ablation}

\begin{figure*}[t]
\centering
\includegraphics[width=\textwidth]{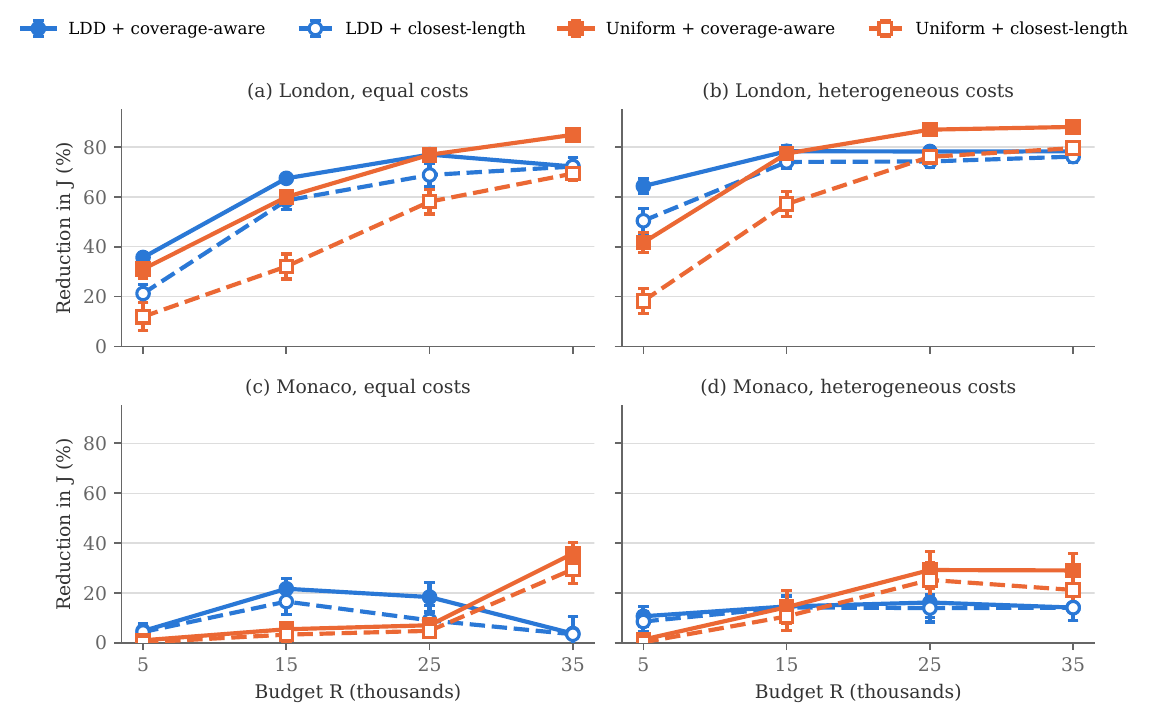}
\caption{Reduction in the coverage penalty $J$ relative to shortest routes, by network, cost setting, budget, deviation rule and selection rule (mean and 95\% interval over 20 paired scenarios, 14 vehicles, clustered placement). Coverage-aware routing is Algorithm~\ref{alg:greedy}.}
\label{fig:ablation}
\end{figure*}

Fig.~\ref{fig:ablation} shows the reduction in $J$ for the four combinations of deviation rule and selection rule; random selection, omitted for clarity, performs close to closest-length selection.

\textit{Coverage-aware routing} improves on closest-length selection in every setting, by up to 27.9 points in London and 9.3 points in Monaco, where an active vehicle has only 1 to 3 candidates on average, against 8 to 20 in London. Under LDD the gain shrinks as more active vehicles become unrealized, because a vehicle without candidates leaves nothing to select. Local search adds at most 1.6 points over the greedy phase.

\textit{Deviation targets.} With coverage-aware routing, LDD improves on uniform deviation at small budgets in all four settings, by up to $22.6\pm3.5$ points (London, heterogeneous costs, $R=5{,}000$). At the largest budget the order reverses in all four, with uniform deviation ahead by $9.8\pm2.5$ to $32.1\pm7.5$ points; the reversal coincides with the detour limits becoming binding for LDD. Supplement~S7 provides additional numerical comparisons.

\subsection{Added Travel and Participation}
\label{sec:eval:incentives}

The equilibrium of Section~\ref{sec:game} concentrates deviation on the vehicles with the lowest effective cost; in the clustered placement, 4 to 5 vehicles are active at every budget, mostly from the isolated group. Because the active set does not change with $R$ (Proposition~\ref{prop:stage1}), their targets grow linearly with the budget and eventually exceed their detour limits: in London every unrealized LDD vehicle has a target above $D_i+\delta$, and with heterogeneous costs on average $1.75$, $2.55$ and $3.05$ of the 4 active vehicles are unrealized at $R=15{,}000$, $25{,}000$ and $35{,}000$. Uniform deviation spreads the same total over all 14 vehicles, so its vehicles drive much more: in London, 70.4\% added travel against 25.6\% for LDD at $R=35{,}000$ with heterogeneous costs.

\begin{figure}[t]
\centering
\includegraphics[width=\columnwidth]{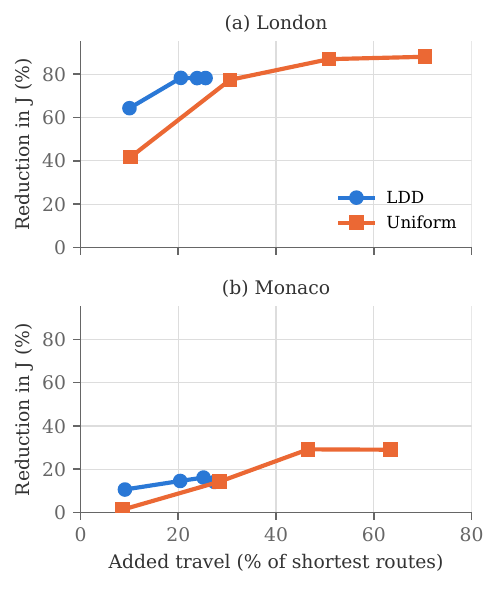}
\caption{Coverage against added travel, heterogeneous costs, coverage-aware routing, clustered placement. Points along each curve are the budgets $R=5$, 15, 25 and $35\times10^{3}$, from left to right.}
\label{fig:travel}
\end{figure}

Fig.~\ref{fig:travel} therefore plots coverage against the travel that produces it. Measured as the mean reduction in $J$ per percentage point of mean added travel, LDD is ahead of uniform deviation at every budget in London with both cost settings and in Monaco with heterogeneous costs, and at three of four budgets in Monaco with equal costs; Supplement~S7 gives the values.

Uniform deviation is also not an equilibrium of the game (Table~\ref{tab:participation}). This comparison concerns equilibrium targets against non-equilibrium targets with the same total deviation; it does not isolate the effect of the diversity weights, which an equilibrium of the same game with unit weights would. Under the payment rule, vehicles with negative utility would prefer their default routes, and compensating them would take up to 88\% of the budget in individual runs with heterogeneous costs. These payments cover the utility shortfalls of the evaluated assignments; they do not make the compensated assignments an equilibrium. Under LDD, negative utilities are rare: 1 of 8,960 vehicle assignments with equal costs and 16 of 8,960 with heterogeneous costs, most of them slightly below zero (minimum $-28.7$, in Monaco), consistent with Corollary~\ref{cor:ir} being only a sufficient condition.

\begin{table}[t]
\centering
\caption{Participation under the payment rule, clustered placement (over all budgets and selection rules)}
\label{tab:participation}
\small
\begin{tabular}{@{}p{0.40\columnwidth}cc@{}}
\hline
 & Equal costs & Heterog.\ costs \\
\hline
\multicolumn{3}{@{}l}{\emph{London}} \\
Vehicles with negative utility, LDD & 0 of 14 & 0.01 of 14 \\
Vehicles with negative utility, uniform & 2.8 of 14 & 8.5 of 14 \\
Compensation, uniform, \% of $R$, mean (max) & 0.9 (4.4) & 41.5 (83.5) \\
\multicolumn{3}{@{}l}{\emph{Monaco}} \\
Vehicles with negative utility, LDD & 0.003 of 14 & 0.04 of 14 \\
Vehicles with negative utility, uniform & 2.7 of 14 & 8.6 of 14 \\
Compensation, uniform, \% of $R$, mean (max) & 1.0 (7.3) & 42.5 (88.1) \\
\hline
\end{tabular}
\end{table}

\subsection{Robustness to Vehicle Placement}
\label{sec:eval:placement}

Table~\ref{tab:placement} repeats the comparison of deviation rules for the three placements with heterogeneous costs. The participation result holds in every case: uniform deviation leaves 8.6 to 9.5 vehicles with negative utility and requires compensation averaging 42\% to 60\% of the budget, and in some balanced and dispersed runs more than the budget itself. The coverage advantage of LDD at small budgets holds for the clustered and balanced placements in both networks and for the dispersed placement in Monaco, but not for the dispersed placement in London, where uniform deviation is ahead by $8.3\pm7.3$ points at $R=5{,}000$ and also uses travel more efficiently at three of four budgets. When the vehicles already spread over the network, the diversity weights vary less (their largest-to-smallest ratio averages 1.6 in the dispersed London scenarios, against 2.4 in the clustered ones), and concentrating the budget on a few vehicles buys less additional coverage than spreading it. At the largest budget uniform deviation is ahead in every placement.

\begin{table}[t]
\centering
\caption{Placement robustness, heterogeneous costs, coverage-aware routing (20 paired scenarios each). $\Delta J$: LDD minus uniform deviation, in points of reduction in $J$; Comp.: compensation needed under uniform deviation, \% of $R$}
\label{tab:placement}
\footnotesize
\setlength{\tabcolsep}{2.5pt}
\begin{tabular}{@{}llccc@{}}
\hline
 & Placement & $\Delta J$, $R{=}5$k & $\Delta J$, $R{=}35$k & Comp. \\
\hline
London & clustered & $+22.6\pm3.5$ & $-9.8\pm2.5$ & 41.9 (83.5) \\
 & balanced & $+21.5\pm2.6$ & $-7.9\pm2.2$ & 59.9 (143.5) \\
 & dispersed & $-8.3\pm7.3$ & $-34.5\pm7.8$ & 47.8 (115.9) \\
Monaco & clustered & $+9.3\pm3.8$ & $-14.9\pm7.0$ & 42.6 (87.7) \\
 & balanced & $+9.8\pm6.1$ & $-10.8\pm9.5$ & 53.4 (166.8) \\
 & dispersed & $+5.8\pm6.2$ & $-21.5\pm9.3$ & 56.1 (115.8) \\
\hline
\end{tabular}
\end{table}

\subsection{Detour Limits and Realizability}
\label{sec:eval:detour}

Detour limits determine how much of the LDD allocation can be realized. In London with heterogeneous costs at $R=35{,}000$, the reduction in $J$ under LDD is $59.9\%$, $78.3\%$ and $82.9\%$ for $D_i=0.5L_i$, $L_i$ and $1.5L_i$, with $3.75$, $3.05$ and $2.40$ unrealized vehicles on average. Uniform deviation is less sensitive, $78.5\%$, $88.1\%$ and $88.2\%$, until its own targets exceed the tightest limit (13.1 unrealized vehicles at $D_i=0.5L_i$). At $R=5{,}000$, where the targets are realizable, LDD stays ahead under every limit (57.9\%, 64.3\% and 65.3\% against 41.7\% for uniform deviation).

Monaco shows a second cause of unrealized vehicles. Its street layout offers few simple routes whose deviation falls within $\delta=100$\,m of a given target: an active vehicle has on average 1.1 candidates under LDD and 2.3 to 2.7 under uniform deviation, and a third of the unrealized LDD vehicles with heterogeneous costs, more than half with equal costs, and almost all unrealized vehicles under uniform deviation have targets below their detour limits. Assumption~\ref{ass:realizable} therefore holds in only 9 of the 160 LDD runs in Monaco, against 58 of 160 in London.

\subsection{Incentive Certificates}
\label{sec:eval:certificates}

\begin{figure*}[t]
\centering
\includegraphics[width=\textwidth]{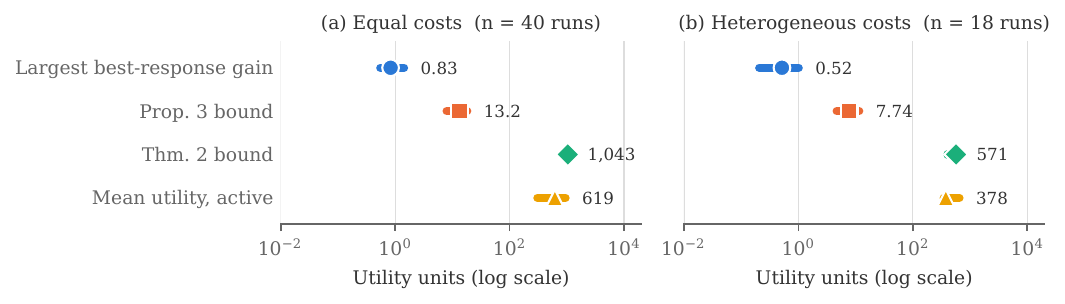}
\caption{Incentive certificates for active vehicles in fully realized London LDD runs with coverage-aware routing: largest best-response gain, an upper bound on the gain from any feasible route change; the bounds of Proposition~\ref{prop:second} and Theorem~\ref{thm:epsne}, and the mean utility of active vehicles (median and interquartile range over runs).}
\label{fig:certificate}
\end{figure*}

Fig.~\ref{fig:certificate} compares the certificates of Section~\ref{sec:incentives} with the largest best-response gain of any active vehicle, in the London LDD runs with coverage-aware routing in which every target is realized. Since this gain upper-bounds what a vehicle can obtain from any feasible route change, small values certify that the routed assignments are close to equilibria: the median largest gain is 0.83 with equal costs and 0.52 with heterogeneous costs, against median utilities of active vehicles of 619 and 378. Theorem~\ref{thm:epsne} certifies every run, but conservatively, with median bounds of 1,043 and 571. The second-order bound of Proposition~\ref{prop:second}, whose condition held for 229 of the 232 active vehicles in these runs, is approximately 15 times the continuous best-response gains (median per-run ratio 14.7 and 15.1) and well below the utilities. For inactive vehicles the best-response gain was at most 0.08, and the median bound \eqref{eq:epsinactive} was 6.6 and 1.9.

In runs with fallback routes, the continuous best-response bounds were substantially larger, and the aggregate-gap condition sometimes failed. In London LDD runs with at least one fallback route, the median largest best-response gain of an active vehicle was 481 with equal costs and 2,445 with heterogeneous costs, and that of an inactive vehicle was 5.2 and 1,590, because the aggregate they face falls far below its equilibrium value. These are upper bounds: whether a vehicle can actually obtain such a gain depends on the routes available to it. With heterogeneous costs the gap condition of Theorem~\ref{thm:epsne} failed in 30 of the 80 London runs and 22 of the 80 Monaco runs, all with fallbacks. Realizability is thus the binding requirement for the incentive guarantees in practice (Section~\ref{sec:discussion}).

\subsection{Numerical Robustness}
\label{sec:eval:numerical}

Supplement~S6 reports checks of grid resolution, the self term and the Stage~2 parameters, with 10 paired scenarios each. Re-scoring the same routes on 50, 100 and 200\,m grids changes the reduction in $J$ by at most 0.8 points, and selecting on a coarser grid leaves the method differences unchanged, so 100\,m cells suffice. Removing the self term from \eqref{eq:diversity}, as in \cite{iccve2022}, makes the weights more uneven and changes the reduction in $J$ under LDD by $-4.5$ to $-2.0$ points in London and $+3.1$ to $+4.0$ points in Monaco. Varying $\rho$, $\delta$, $K$ and $\eta$ changes it by at most 3.6 points and leaves the number of unrealized vehicles nearly unchanged, so in London the detour limits, not candidate generation, govern realizability.

\subsection{Runtime}
\label{sec:eval:runtime}

Stage~1 takes less than 0.5\,ms. Preparing candidates for all 14 vehicles, including their coverage values, takes a median of 24\,s in London and 2\,s in Monaco in our single-threaded Python implementation, dominated by path reconstruction, Fr\'echet pruning and distance computations; coverage-aware selection takes a median of 14\,ms in London (at most 0.11\,s) and 1.3\,ms in Monaco. Candidate preparation is independent across vehicles and can be parallelized.

%% file: sec_discussion.tex
\section{Discussion}
\label{sec:discussion}

\textit{What the results attribute to each stage.} The participation benefit of LDD over uniform deviation comes from equilibrium allocation, which gives every vehicle nonnegative utility with or without diversity weights, so it is not evidence for the weights themselves. The weights matter for coverage when they differ, that is, when some vehicles are isolated; with dispersed vehicles they are nearly equal, and the equilibrium concentrates deviation on the cheapest vehicles rather than on useful locations. Stage~1 targets equilibrium and participation; coverage enters only in Stage~2.

\textit{Detour limits.} By Proposition~\ref{prop:stage1}(a), the active set does not change with the budget and the targets grow in proportion to it, so beyond some budget the active vehicles cannot drive their targets. The continuous equilibrium assigns zero deviation to inactive vehicles and does not reallocate targets when active vehicles reach their detour limits; in London this happened from $R=15{,}000$, which explains why uniform deviation covered more at large budgets. The same property gives a safeguard: the largest budget whose targets respect every limit,
\begin{equation}
\bar R=\min_{i\in H}\frac{D_i}{t_i^*(1)},
\label{eq:rbar}
\end{equation}
can be computed before a budget is announced, and a larger budget can be spent over several rounds or on more vehicles. Adding the limits to the game, with $t_i\in[0,D_i]$, is the principled fix, but existence of an equilibrium would need its own argument, because the payment \eqref{eq:payment} is discontinuous at the profile in which no vehicle deviates, and the closed form of Theorem~\ref{thm:stage1} and the budget-independent active set would no longer hold. We leave this detour-aware equilibrium to future work.

\textit{Realizability.} The a priori guarantees in terms of $\delta$ require Assumption~\ref{ass:realizable}. The actual-gap certificate of Theorem~\ref{thm:epsne} also applies to fallback assignments whenever its aggregate-gap condition holds, but realizability, rather than the size of the bounds, decided whether the certificates were informative: with fallback routes, the best-response bounds reached thousands of utility units and did not certify small incentive losses. In London every unrealized vehicle had a target beyond its detour limit, which $\bar R$ prevents; in Monaco many targets were within the limits, but the via-route generator found few simple routes of the required length. A larger $\delta$ realizes more targets but weakens Theorem~\ref{thm:epsne} linearly. A generator aimed at the target length, for example with two via junctions, would address the cause; exact solutions are hard in general, since deciding whether a simple path of a prescribed length exists contains the Hamiltonian path problem, but heuristics suffice. Because the certificate is computable from the realized gaps, a platform can check it before issuing recommendations and adjust the budget or targets if it fails.

\textit{Modeling assumptions and scale.} Fixed diversity weights keep the route game reducible to a deviation game; updating them would make each utility depend on other vehicles' routes. Costs are proportional to added distance, ignoring travel time and congestion; costs and detour limits are assumed to be reported truthfully, and eliciting them truthfully is a separate problem; and the coverage measure is a static squared-distance penalty without data freshness or location-dependent value. These can be relaxed in the routing stage, but relaxing them in the game requires new equilibrium analysis. The fleets in our experiments are small, as in our earlier work; candidate preparation dominated the runtime but is independent across vehicles and parallelizes directly, and periodic recomputation as trips start and end would also limit the effect of fixed weights.

%% file: sec_conclusion.tex
\section{Conclusion}
\label{sec:conclusion}

We extended location-diversity incentives for vehicular crowdsensing with coverage-aware route selection, separating how far each vehicle deviates, set by a diversity-weighted game, from which route realizes the deviation, chosen to improve collective coverage, and we bounded the incentive loss of the routed assignment. On the London and Monaco road networks, the router attained on average 99.4\% and 98.9\% of the optimal reduction in the coverage penalty on small instances and improved on closest-length selection in every setting. The routed LDD assignments yielded nonnegative utility in nearly all evaluated vehicle assignments, whereas with heterogeneous costs the uniform alternative left most vehicles worse off and would need compensation averaging 42\% of the budget. The LDD targets also bought more coverage per unit of added travel in most settings; their coverage advantage at small budgets is strongest for clustered vehicles and is not consistent across dispersed placements, and it reverses at large budgets, where the detour limits bind.

A detour-constrained deviation game could reduce target-limit violations, while improved candidate generation could reduce failures caused by the current search procedure; neither can create suitable routes where the road network has none. Time-dependent coverage, congestion-aware costs and truthful elicitation of costs and detour limits would bring the mechanism closer to deployment.

%% file: refs.bib
@inproceedings{iccve2022,
  author    = {Chintakunta, Harish and Wang, Xin and Jaimes, Luis G.},
  title     = {Improving Sensing Coverage in Vehicular Crowdsensing using Location Diversity},
  booktitle = {Proc. Int. Conf. Connected Veh. Expo (ICCVE)},
  year      = {2022},
  doi       = {10.1109/ICCVE52871.2022.9742961},
  pages = {1--6}
}

@inproceedings{he2015,
  author    = {He, Zongjian and Cao, Jiannong and Liu, Xuefeng},
  title     = {High Quality Participant Recruitment in Vehicle-Based Crowdsourcing Using Predictable Mobility},
  booktitle = {Proc. IEEE Conf. Comput. Commun. (INFOCOM)},
  year      = {2015},
  pages = {2542--2550},
  doi = {10.1109/INFOCOM.2015.7218644}
}

@article{wang2017,
  author  = {Wang, Xiumin and Wu, Weiwei and Qi, Deyu},
  title   = {Mobility-Aware Participant Recruitment for Vehicle-Based Mobile Crowdsensing},
  journal = {IEEE Trans. Veh. Technol.},
  volume  = {67},
  number  = {5},
  year    = {2018},
  pages = {4415--4426},
  doi = {10.1109/TVT.2017.2787750}
}

@inproceedings{kong2017,
  author    = {Kong, Kaiwei and Xu, Jian and Xu, Ming and Tu, Liming and Wu, Yang and Chen, Zhi},
  title     = {Trajectory Query Based on Trajectory Segments with Activities},
  booktitle = {Proc. 3rd ACM SIGSPATIAL Workshop Smart Cities Urban Analytics},
  year      = {2017},
  pages = {2:1--2:8}
}

@article{chen2017,
  author  = {Chen, Yueyue and Lv, Pin and Guo, Deke and Zhou, Tongqing and Xu, Ming},
  title   = {Trajectory Segment Selection with Limited Budget in Mobile Crowd Sensing},
  journal = {Pervasive Mobile Comput.},
  volume  = {40},
  year    = {2017},
  pages = {123--138},
  doi = {10.1016/j.pmcj.2017.06.010}
}

@article{chakeri2021,
  author  = {Chakeri, Alireza and Wang, Xin and Goss, Quentin and Akbas, M. Ilhan and Jaimes, Luis G.},
  title   = {A Platform-Based Incentive Mechanism for Autonomous Vehicle Crowdsensing},
  journal = {IEEE Open J. Intell. Transp. Syst.},
  volume  = {2},
  year    = {2021},
  doi     = {10.1109/OJITS.2021.3056925},
  pages = {13--23}
}

@inproceedings{chintakunta2021,
  author    = {Chintakunta, Harish and Kahr, Janar and Jaimes, Luis},
  title     = {A Vehicular Crowdsensing Incentive Mechanism for Temporal Coverage},
  booktitle = {Proc. IEEE Consum. Commun. Netw. Conf. (CCNC)},
  year      = {2021},
  doi = {10.1109/CCNC49032.2021.9369634},
  pages = {1--8}
}

@article{elwakeel2018,
  author  = {El-Wakeel, Amr S. and Li, Jin and Noureldin, Aboelmagd and Hassanein, Hossam S. and Zorba, Nizar},
  title   = {Towards a Practical Crowdsensing System for Road Surface Conditions Monitoring},
  journal = {IEEE Internet Things J.},
  year    = {2018},
  volume = {5},
  number = {6},
  pages = {4672--4685},
  doi = {10.1109/JIOT.2018.2807408}
}

@inproceedings{wang2016speed,
  author    = {Wang, Cheng and Zhang, Zhenzhen and Shao, Lu and Zhou, MengChu},
  title     = {Estimating Travel Speed via Sparse Vehicular Crowdsensing Data},
  booktitle = {Proc. IEEE 3rd World Forum Internet Things (WF-IoT)},
  year      = {2016},
  pages = {643--648},
  doi = {10.1109/WF-IoT.2016.7845432}
}

@inproceedings{sheikh2014map,
  author    = {Sheikh, Shuja Jamil and Basalamah, Anas and Aly, Heba and Youssef, Moustafa},
  title     = {Demonstrating {Map++}: A Crowd-Sensing System for Automatic Map Semantics Identification},
  booktitle = {Proc. IEEE Int. Conf. Sensing, Commun., Netw. (SECON)},
  year      = {2014},
  pages = {152--154},
  doi = {10.1109/SAHCN.2014.6990340}
}

@inproceedings{yang2012,
  author    = {Yang, Dejun and Xue, Guoliang and Fang, Xi and Tang, Jian},
  title     = {Crowdsourcing to Smartphones: Incentive Mechanism Design for Mobile Phone Sensing},
  booktitle = {Proc. 18th Annu. Int. Conf. Mobile Comput. Netw. (MobiCom)},
  year      = {2012},
  pages = {173--184}
}

@article{fisher1978,
  author  = {Fisher, Marshall L. and Nemhauser, George L. and Wolsey, Laurence A.},
  title   = {An Analysis of Approximations for Maximizing Submodular Set Functions {II}},
  journal = {Math. Program. Study},
  volume  = {8},
  year    = {1978},
  pages   = {73--87}
}

@article{nemhauser1978,
  author  = {Nemhauser, George L. and Wolsey, Laurence A. and Fisher, Marshall L.},
  title   = {An Analysis of Approximations for Maximizing Submodular Set Functions {I}},
  journal = {Math. Program.},
  volume  = {14},
  year    = {1978},
  pages   = {265--294}
}

@article{calinescu2011,
  author  = {C{\u{a}}linescu, Gruia and Chekuri, Chandra and P{\'a}l, Martin and Vondr{\'a}k, Jan},
  title   = {Maximizing a Monotone Submodular Function Subject to a Matroid Constraint},
  journal = {SIAM J. Comput.},
  volume  = {40},
  number  = {6},
  year    = {2011},
  pages   = {1740--1766}
}

@incollection{minoux1978,
  author    = {Minoux, Michel},
  title     = {Accelerated Greedy Algorithms for Maximizing Submodular Set Functions},
  booktitle = {Optimization Techniques},
  series    = {Lecture Notes in Control and Information Sciences},
  volume    = {7},
  publisher = {Springer},
  year      = {1978},
  pages   = {234--243}
}

@article{dimartino2022routr,
  author  = {Di Martino, Sergio and Starace, Luigi Libero Lucio},
  title   = {Towards Uniform Urban Map Coverage in Vehicular Crowd-Sensing: A Decentralized Incentivization Solution},
  journal = {IEEE Open J. Intell. Transp. Syst.},
  volume  = {3},
  year    = {2022},
  doi     = {10.1109/OJITS.2022.3211540},
  pages = {695--708}
}

@article{chakeri2022market,
  author  = {Chakeri, Alireza and Wang, Xin and Jaimes, Luis G.},
  title   = {A Vehicular Crowdsensing Market for {AVs}},
  journal = {IEEE Open J. Intell. Transp. Syst.},
  volume  = {3},
  year    = {2022},
  doi     = {10.1109/OJITS.2022.3162756},
  pages = {278--287}
}

@article{jaimes2024sensenow,
  author  = {Jaimes, Luis G. and Chintakunta, Harish and Abedin, Paniz},
  title   = {{SenseNow}: A Time-Dependent Incentive Approach for Vehicular Crowdsensing},
  journal = {IEEE Open J. Intell. Transp. Syst.},
  volume  = {5},
  year    = {2024},
  doi     = {10.1109/OJITS.2024.3411525},
  pages = {307--321}
}

@inproceedings{wang2021navigation,
  author    = {Wang, En and Luan, Dongming and Yang, Yongjian and Wang, Zihe and Dong, Pengmin and Li, Dawei and Liu, Wenbin and Wu, Jie},
  title     = {Distributed Game-Theoretical Route Navigation for Vehicular Crowdsensing},
  booktitle = {Proc. 50th Int. Conf. Parallel Process. (ICPP)},
  year      = {2021},
  doi       = {10.1145/3472456.3472498},
  pages = {67:1--67:11}
}

@article{ji2016discretized,
  author  = {Ji, Shiyu and Chen, Tingting},
  title   = {Incentive Mechanisms for Discretized Mobile Crowdsensings},
  journal = {IEEE Trans. Wireless Commun.},
  volume  = {15},
  number  = {1},
  year    = {2016},
  doi     = {10.1109/TWC.2015.2468724},
  pages = {146--161}
}

@article{xu2020ilocus,
  author  = {Xu, Susu and Chen, Xinlei and Pi, Xidong and Joe-Wong, Carlee and Zhang, Pei and Noh, Hae Young},
  title   = {{iLOCuS}: Incentivizing Vehicle Mobility to Optimize Sensing Distribution in Crowd Sensing},
  journal = {IEEE Trans. Mobile Comput.},
  volume  = {19},
  number  = {8},
  year    = {2020},
  doi     = {10.1109/TMC.2019.2915838},
  pages = {1831--1847}
}

@misc{guo2023rerouting,
  author       = {Guo, Shuocheng and Qian, Xinwu},
  title        = {Optimal Drive-by Sensing in Urban Road Networks with Large-scale Ridesourcing Vehicles},
  howpublished = {arXiv preprint arXiv:2207.11285 (v2)},
  year         = {2023}
}

@inproceedings{masutani2015,
  author    = {Masutani, Osamu},
  title     = {A Sensing Coverage Analysis of a Route Control Method for Vehicular Crowd Sensing},
  booktitle = {Proc. IEEE Int. Conf. Pervasive Comput. Commun. Workshops (PerCom Workshops)},
  year      = {2015},
  pages = {396--401},
  doi = {10.1109/PERCOMW.2015.7134070}
}

@inproceedings{asprone2018,
  author    = {Asprone, Dario and Di Martino, Sergio and Festa, Paola},
  title     = {Improving Sensing Coverage of Probe Vehicles with Probabilistic Routing},
  booktitle = {Web and Wireless Geographical Information Systems (W2GIS)},
  series    = {Lecture Notes in Computer Science},
  volume    = {10819},
  publisher = {Springer},
  year      = {2018},
  pages = {1--10},
  doi = {10.1007/978-3-319-90053-7_1}
}

@article{asprone2021,
  author  = {Asprone, Dario and Di Martino, Sergio and Festa, Paola and Starace, Luigi Libero Lucio},
  title   = {Vehicular Crowd-Sensing: A Parametric Routing Algorithm to Increase Spatio-Temporal Road Network Coverage},
  journal = {Int. J. Geogr. Inf. Sci.},
  volume  = {35},
  number  = {9},
  year    = {2021},
  pages = {1876--1904},
  doi = {10.1080/13658816.2021.1893737}
}

@article{zhang2015quality,
  author  = {Zhang, Maotian and Yang, Panlong and Tian, Chang and Tang, Shaojie and Gao, Xiaofeng and Wang, Baowei and Xiao, Fu},
  title   = {Quality-Aware Sensing Coverage in Budget-Constrained Mobile Crowdsensing Networks},
  journal = {IEEE Trans. Veh. Technol.},
  volume  = {65},
  number  = {9},
  pages   = {7698--7707},
  year    = {2016},
  doi     = {10.1109/TVT.2015.2490679}
}

@article{han2024mixed,
  author  = {Han, Ke and Ji, Wen and Nie, Yu (Marco) and Li, Zhexian and Liu, Shenglin},
  title   = {Exploring the Sensing Power of Mixed Vehicle Fleets},
  journal = {Transp. Res. B, Methodol.},
  volume  = {190},
  year    = {2024},
  note = {Art. no. 103066},
  doi = {10.1016/j.trb.2024.103066}
}

@article{dai2023bus,
  author  = {Dai, Zhuang and Han, Ke},
  title   = {Exploring the Drive-by Sensing Power of Bus Fleet through Active Scheduling},
  journal = {Transp. Res. E, Logist. Transp. Rev.},
  volume  = {171},
  year    = {2023},
  note = {Art. no. 103029},
  doi = {10.1016/j.tre.2023.103029}
}

@article{ji2023survey,
  author  = {Ji, Wen and Han, Ke and Liu, Tao},
  title   = {A Survey of Urban Drive-by Sensing: An Optimization Perspective},
  journal = {Sustain. Cities Soc.},
  year    = {2023},
  note    = {Art. no. 104874},
  volume = {99}
}

@article{feng2025nonproprietary,
  author  = {Feng, Zhirui and Yu, Yantao and Liu, Guojin and Jiang, Yang and Huang, TianCong},
  title   = {Incentive Mechanisms for Non-Proprietary Vehicles in Vehicular Crowdsensing with Budget Constraints},
  journal = {Comput. Commun.},
  volume  = {235},
  note    = {Art. no. 108083},
  year    = {2025},
  doi     = {10.1016/j.comcom.2025.108083}
}

@article{zhang2026bepim,
  author  = {Zhang, Jiamin and Shuai, Lisha and Dong, Jiuling and Dong, Gaoya and Yang, Xiaolong and Long, Keping},
  title   = {{BEP-IM}: A Vehicular Crowdsensing Incentive Mechanism to Drive Sustained Spatial Coverage and Proactive Sensing Shaping},
  journal = {Entropy},
  volume  = {28},
  number  = {5},
  note    = {Art. no. 499},
  year    = {2026},
  doi     = {10.3390/e28050499}
}

@article{krajzewicz2012sumo,
  author  = {Krajzewicz, Daniel and Erdmann, Jakob and Behrisch, Michael and Bieker, Laura},
  title   = {Recent Development and Applications of {SUMO} - {Simulation of Urban MObility}},
  journal = {Int. J. Adv. Syst. Meas.},
  volume  = {5},
  number  = {3, 4},
  year    = {2012},
  pages = {128--138}
}

@inproceedings{codeca2017most,
  author    = {Codec{\'a}, Lara and H{\"a}rri, J{\'e}r{\^o}me},
  title     = {Towards Multimodal Mobility Simulation of {C-ITS}: The {Monaco} {SUMO} Traffic Scenario},
  booktitle = {Proc. IEEE Veh. Netw. Conf. (VNC)},
  address   = {Turin, Italy},
  year      = {2017},
  doi       = {10.1109/VNC.2017.8275627},
  pages   = {97--100}
}
